\documentclass[11pt]{article}

\date{}

\usepackage[utf8]{inputenc} 
\usepackage[T1]{fontenc}    
\usepackage[colorlinks=true,linkcolor=blue,allcolors=blue]{hyperref}       
\usepackage{url}            
\usepackage{booktabs}       
\usepackage{amsfonts}       
\usepackage{nicefrac}       
\usepackage{microtype,nicefrac}      
\usepackage{xspace}
\usepackage{natbib}

\usepackage{comment}
\usepackage{amsmath}
\usepackage{graphicx}
\usepackage{rotating}

\usepackage{amssymb}
\usepackage{autobreak}
\usepackage{mathtools}
\usepackage[dvipsnames]{xcolor}
\usepackage{bbm}
\usepackage[ruled,vlined, linesnumbered]{algorithm2e}
\usepackage{algorithmic}
\usepackage[parfill]{parskip}

\usepackage[toc]{appendix}
\usepackage{minitoc}

\makeatletter

\renewcommand{\L}{\mathcal L}

\newtheorem{proof*}{Proof}[section]

\title{Seismic wave propagation and inversion with Neural Operators}

\author{%
  Yan Yang$^{1*}$, Angela F. Gao$^{2*}$, Jorge C. Castellanos$^{1}$\footnote{These authors contributed equally to this work.}, Zachary E. Ross$^1$, \\
  Kamyar Azizzadenesheli$^3$, and Robert W. Clayton$^1$ \\ \\
  {\small ${}^1$Seismological Laboratory, California Institute of Technology, Pasadena, CA}\\
  {\small ${}^2$Dept. of Computing and Mathematical Sciences, California Institute of Technology, Pasadena, CA}\\
  {\small ${}^3$Dept. of Computer Science, Purdue University, West Lafayette, IN}\\
}

\begin{document}
\doparttoc 
\faketableofcontents 

\maketitle

\begin{abstract}
Seismic wave propagation forms the basis for most aspects of seismological research, yet solving the wave equation is a major computational burden that inhibits the progress of research. This is exacerbated by the fact that new simulations must be performed when the velocity structure or source location is perturbed. Here, we explore a prototype framework for learning general solutions using a recently developed machine learning paradigm called Neural Operator. A trained Neural Operator can compute a solution in negligible time for any velocity structure or source location. We develop a scheme to train Neural Operators on an ensemble of simulations performed with random velocity models and source locations. As Neural Operators are grid-free, it is possible to evaluate solutions on higher resolution velocity models than trained on, providing additional computational efficiency. We illustrate the method with the 2D acoustic wave equation and demonstrate the method's applicability to seismic tomography, using reverse mode automatic differentiation to compute gradients of the wavefield with respect to the velocity structure. The developed procedure is nearly an order of magnitude faster than using conventional numerical methods for full waveform inversion.
\end{abstract}

\section{Introduction}
The simulation of seismic wave propagation through Earth's interior underlies most aspects of seismological research, from the simulation of strong ground shaking due to large earthquakes \citep{rodgers_broadband_2019,graves_kinematic_2016}, to imaging the subsurface velocity structure \citep{tape_adjoint_2009,virieux_overview_2009,fichtner_full_2009,lee_full-3-d_2014,gebraad_bayesian_2020}, to derivation of earthquake source properties \citep{duputel_iquique_2015,ye_rupture_2016,wang_moving_2020}. The compute costs associated with these wavefield simulations are substantial, and for reasons of computational efficiency, 1D models are often used, even when better 3D velocity models are available. As a result, seismic wave simulations are often the limiting factor in the pace of geophysical research.

Recently, deep learning approaches have been explored with the goal of solving various geophysical partial differential equations \citep{smith_eikonet_2020,moseley_solving_2020,moseley_deep_2020,moseley_finite_2021}. Beyond the goal of accelerating compute capabilities, such physics-informed neural networks may offer other advantages such as grid-independence, low memory overhead, differentiability, and on-demand solutions. These properties can then result in deep learning being used to solve geophysical inverse problems \citep{zhu_integrating_2020,zhang_deep-learning_2021,xiao_deep-learning-based_2021,smith_hyposvi_2021}, as a wider selection of algorithms and frameworks then are available for use, such as approximate Bayesian inference techniques like variational inference. 


One of the major challenges associated with wave propagation is that a new simulation must be performed whenever the properties of the source or velocity structure are perturbed in some way. This alone substantially increases the necessary compute costs, making some problems prohibitively expensive even if they are mathematically or physically tractable. For the most part, these limitations have been accepted as an inevitable part of seismology, but now physics-informed machine learning approaches have started to offer some pathways for moving beyond this issue. For example, \cite{smith_eikonet_2020} use a deep neural network to solve the Eikonal equation for any source-receiver pair by taking these locations as input. This then can be exploited for hypocenter inversion by allowing for gradients of the travel time field to be computed with respect to the source location \citep{smith_hyposvi_2021}. However, these models are relatively inefficient to train and even then, are only able to learn approximate solution operators to the differential equations.

The aforementioned limitations may seem surprising, but result from a basic attribute of neural networks that in fact makes them ill-suited for solving differential equations. Specifically, neural networks are designed for learning maps between two finite dimensional spaces, while learning a general solution operator for a differential equation requires the ability to map between two infinite dimensional spaces (i.e. function spaces). A paradigm for learning maps between function spaces was recently developed \citep{li_neural_2020,li_multipole_2020,li_fourier_2021}, and has been termed Neural Operator. The general idea behind these models is that they have shared parameters over all possible functions describing the initial conditions, which allows them operate on functions, even when the inputs are a numerically discretized representation of them.

Here, we explore the potential of Neural Operators in improving seismic wave propagation and inversion. We develop a prototype framework for training Neural Operators on the 2D acoustic wave equation and show that this approach provides a suite of tantalizing new advantages over conventional numerical methods for seismic wave propagation. This study provides a proof of concept of this technology and its application to seismology.

\section{Preliminaries}

For a given function $A$ and a Green's function $G$, let $U$ denote the solution to a linear partial differential equation, i.e., the solution operator,

\begin{align*}
    U(x) = (\L A)(x) =\int G(x,y)A(y)dy
\end{align*}

where $\L$ is the corresponding linear operator. For example, suppose that the PDE to be solved is the acoustic wave equation; then $A$ could describe the velocity structure as well as the initial conditions. Neural operator generalizes this formulation to the nonlinear setting where a set of linear operators are sequentially applied to construct a general nonlinear solution operator. In its basic form, an $\ell$-layered neural operator is constructed as follows; 

\begin{align*}
    U(x)=\L_\ell\left(\sigma(\L_{\ell-1} \ldots \sigma\left(\L_1V\right)\ldots)\right)(x) 
\end{align*}
where $\L_i$ is such that for any function $V$, we have,

\begin{equation}
    (\L_i V)(x) = W_i(x)+\int K_i(x,y)V(y)dy
\end{equation}

Under this framework, $W_i(x)$ and $K_i(x,y)$ constitute the learnable components of the Neural Operator and allow for a solution to be produced for any prescribed function $A$. In a limited sense, Neural Operators can be viewed as generalized Green's functions.

\section{Methods}\label{Sec:SV}

\begin{figure}[!ht]
    \centering
    \includegraphics[width=0.90\textwidth]{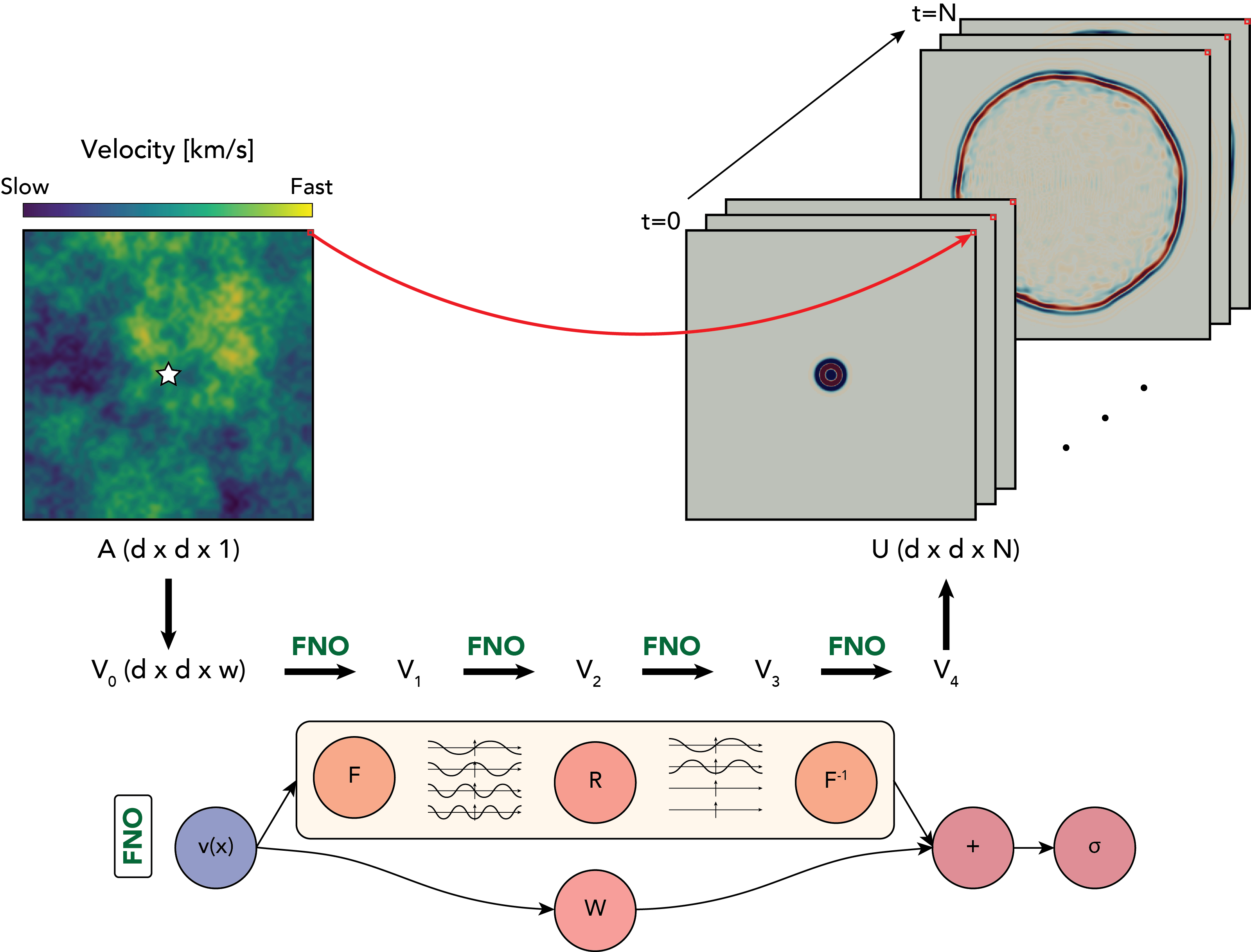}
    \caption{Our approach applying FNO to the 2D acoustic wave equation. The inputs to the FNO model are the velocity model with dimensions $d \times d \times 1$ and the source location, indicated by the white star. The input velocity model is lifted to a higher dimensional space with size $d \times d  \times w$ using a neural network. Then, we apply four Fourier operator layers, and finally project back to the target wavefield dimensions of $d \times d \times N$ using a neural network. The bottom panel shows details of the Fourier layer architecture: we define $v$ to be the input. On top: We apply a Fourier transform F to $v$, then apply a linear transformation R to the lower Fourier modes, filtering out higher modes. Then we apply an inverse Fourier transform F$^{-1}$. On the bottom: we apply a local linear transform W to $v$.}
    \label{fig:2D_diagram}
\end{figure}

We designed a framework that applies Neural Operators to the 2D acoustic wave equation. The basic idea for this procedure is outlined schematically in Figure \ref{fig:2D_diagram}. A specific type of Neural Operator, called a Fourier Neural Operator \citep[FNO;][]{li_fourier_2021} receives a velocity model specified on an arbitrary, possibly irregular mesh, along with the coordinates of a point source. One of the main features of FNO is that the heavy calculations are performed in the frequency domain, which allows the convolutions in eq. 1 to be rapidly computed. The output of the FNO is the complete wavefield solution, which can be queried anywhere within the medium, regardless of whether the points lie on the input mesh.

The most basic component of the FNO is a Fourier block (Fig. \ref{fig:2D_diagram}), which first transforms an input function ($V$) to the Fourier domain. In the first layer of the network, $V$ is equal to the initial conditions, $A$. Then, a kernel ($K_i$) is computed specifically for this function and is truncated at low order, before performing the integration via multiplication. Finally, the result is transformed back and a non-linear activation function is applied, which concludes the Fourier block. For this study, the architecture of the FNO contains 4 sequential Fourier blocks and applies a ReLU activation to the output of each (Fig. \ref{fig:2D_diagram}). We note that the truncation of the Fourier modes is performed on the function values after lifting them to a higher dimensional space, rather than the raw input function, so that this does not lead to compression.

We constructed a training dataset of simulations to learn from by first generating random velocity models. We set up a 64$\times$64 grid with 0.16 km spacing to define the velocity model. Then, we created 5,000 random velocity models by sampling random fields having a von Karman covariance function with the following parameters: Hurst exponent $\kappa =  0.5$, correlation length $\textbf{a} = [a_x, a_y]=[0.16 \text{km},0.16 \text{km}]$, and $\varepsilon =  0.1$, $\mu = 3 \text{km/s}$, and $\sigma = 0.15 \text{km}$. Then, for each of these velocity models, we apply a Ricker wavelet source at a random point, and solve the acoustic wave equation using a spectral-element method (SEM) \citep{afanasiev_modular_2019}. It should be noted that there is a source grid used since this is a requirement of the spectral-element method. As Gaussian random fields can represent all continuous functions, the purpose of these steps is to create a suite of simulations that span a range of possible conditions. We show later that they can even well-approximate strongly discontinuous velocity models. An example velocity model and simulation is shown in Figure \ref{fig:2D_diagram}. Applying the aforementioned procedure results in a training dataset of 5,000 data samples, each of which is a different simulation.

Given the simulation dataset, we can proceed to train a FNO model in a supervised capacity, where the goal is to learn a model that can reliably output a solution to the wave equation for arbitrary input conditions. The training is performed using batch gradient descent, where the parameters of the FNO are updated to minimize the error against the "true" spectral-element solutions. A mean-square error loss function is used. We use a batch size of 30 simulations and train the model for a maximum of 300 epochs. We use all but 200 of the simulations for training, and set aside the remainder for cross-validation of the model. The time required to train the model from scratch is approximately 18 hours using 6 NVIDIA Tesla V100 GPUs. 

\section{Experiments}\label{Sec:Exp}

\subsubsection*{Initial wavefield demonstration}
\begin{figure}[!ht]
    \centering
    \includegraphics[width=0.95\textwidth]{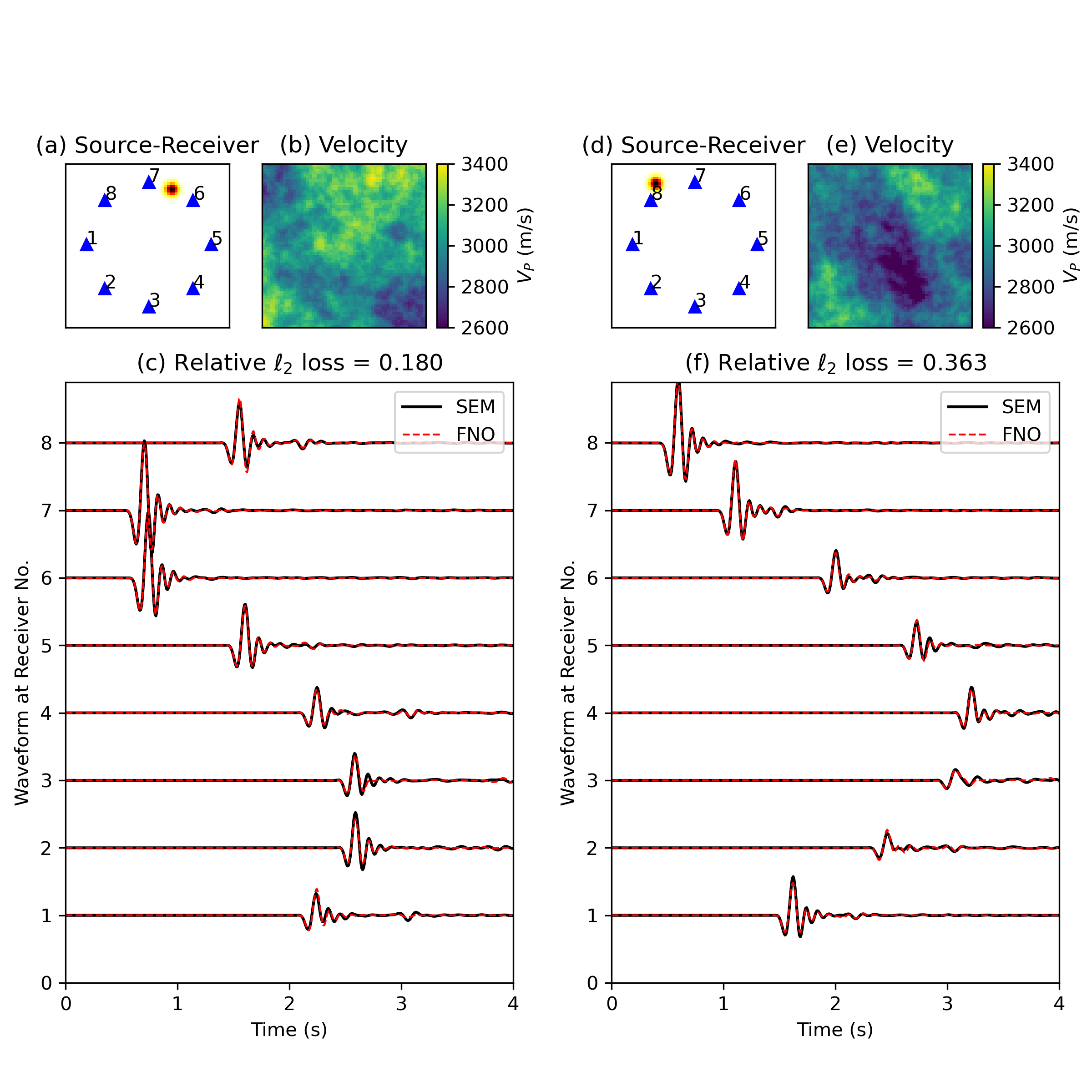}
    \caption{Examples of two validation wavefield simulations from the trained FNO model. (a) the source-receiver locations with receivers in blue and source in red; (b) the velocity structure;  (c) waveforms simulated with SEM (black) and FNO (red). (d-f) same as (a-c), but for a different velocity model. The relative $\ell_2$ loss of the two examples are 0.180 and 0.363, respectively, which are representative of the entire validation data set with an average $\ell_2$ loss of 0.273. We demonstrate that the FNO simulation results are able to capture both the major arrivals as well as some reflections.}
    \label{fig:example_wavefields}
\end{figure}

Figure \ref{fig:example_wavefields} shows two example wavefields corresponding to two different velocity models, each with a different source. The spectral element solution is shown alongside the wavefield predicted by the FNO for the 8 different receivers (blue triangles). For these examples, the input velocity model is 64$\times$64. The relative $\ell_2$ loss of the FNO wavefields are 0.180 and 0.363. These examples are representative of the entire validation dataset, which has a loss of 0.273 relative to the spectral element solutions. 

\subsubsection*{The number of simulations needed for training}
\begin{figure}[!ht]
    \centering
    \includegraphics[width=0.90\textwidth]{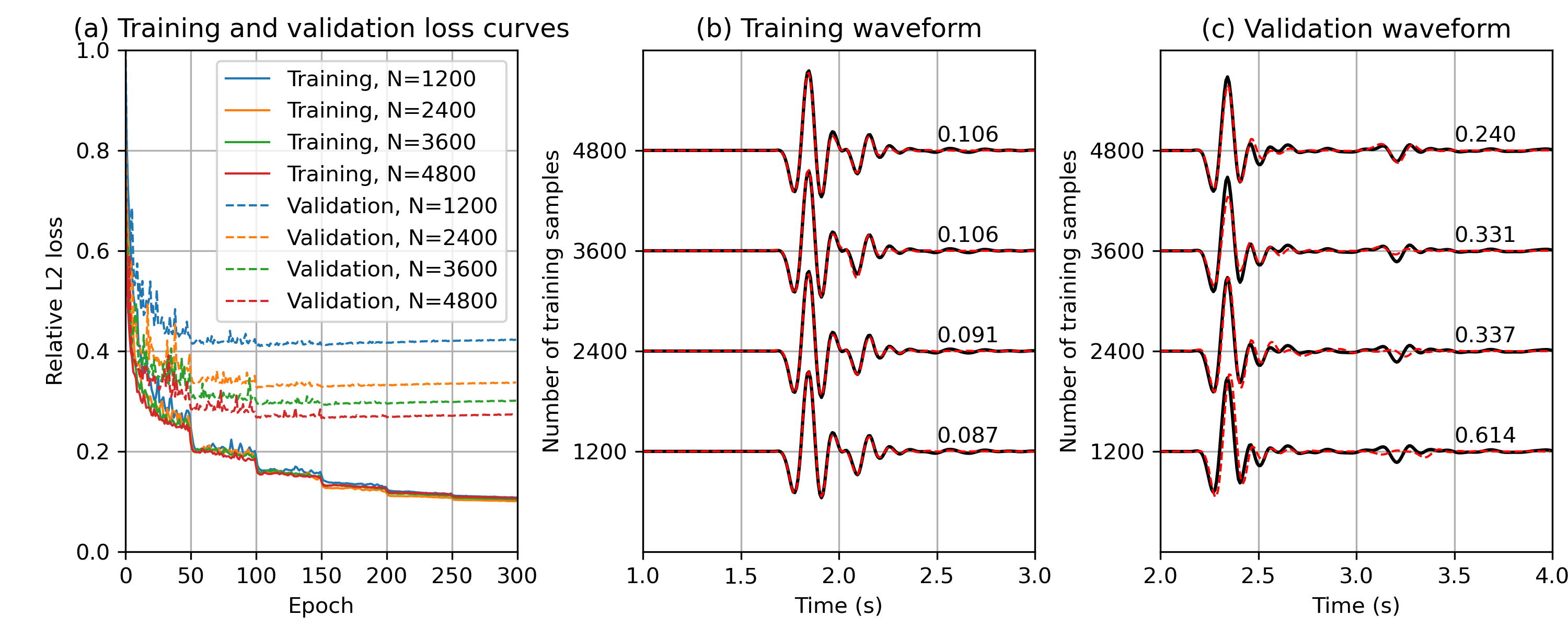}
    \caption{Model performance as a function of the number of training samples. (a) Training and validation loss curves as a function of different numbers of training samples. (b) Example waveform fitting of a single training example from models trained with varying number of training examples. (c) Example waveform fitting of a single validation example from models trained with varying number of training examples. The numbers to the right of each waveform shows the relative $\ell_2$ misfit. This shows that the model trained on 4800 samples is able to capture the reflections, whereas the model trained on smaller number of samples does not generalize to reflections in the validation example.}
    \label{fig:number_samples}
\end{figure}

Once fully trained, the FNO can evaluate a new solution in a fraction of a second, and thus the time to train the FNO will be the vast majority of the needed compute time. A primary concern about the computational demands of the FNO approach is therefore the number of simulations needed for training. Here, we examine how the number of training simulations influences the accuracy of the solution. We create a series of tests in which the number of training simulations is varied from 800 at the fewest, to 4800 at the most. The results are shown in Figure \ref{fig:number_samples} where we show the FNO wavefield predictions for each dataset. Even with 1200 training samples, there is no indication that there is overfitting since the training waveform error is similar across different models (Figure \ref{fig:number_samples}a,b). Training using just 1200 simulations is able to predict the major arrival. Increasing number of training samples provides a better fit of the reflections (e.g. 3.2 sec in Figure \ref{fig:number_samples}c). 

\subsubsection*{Generalization to arbitrary velocity models}
\begin{figure}[!ht]
    \centering
    \includegraphics[width=0.90\textwidth]{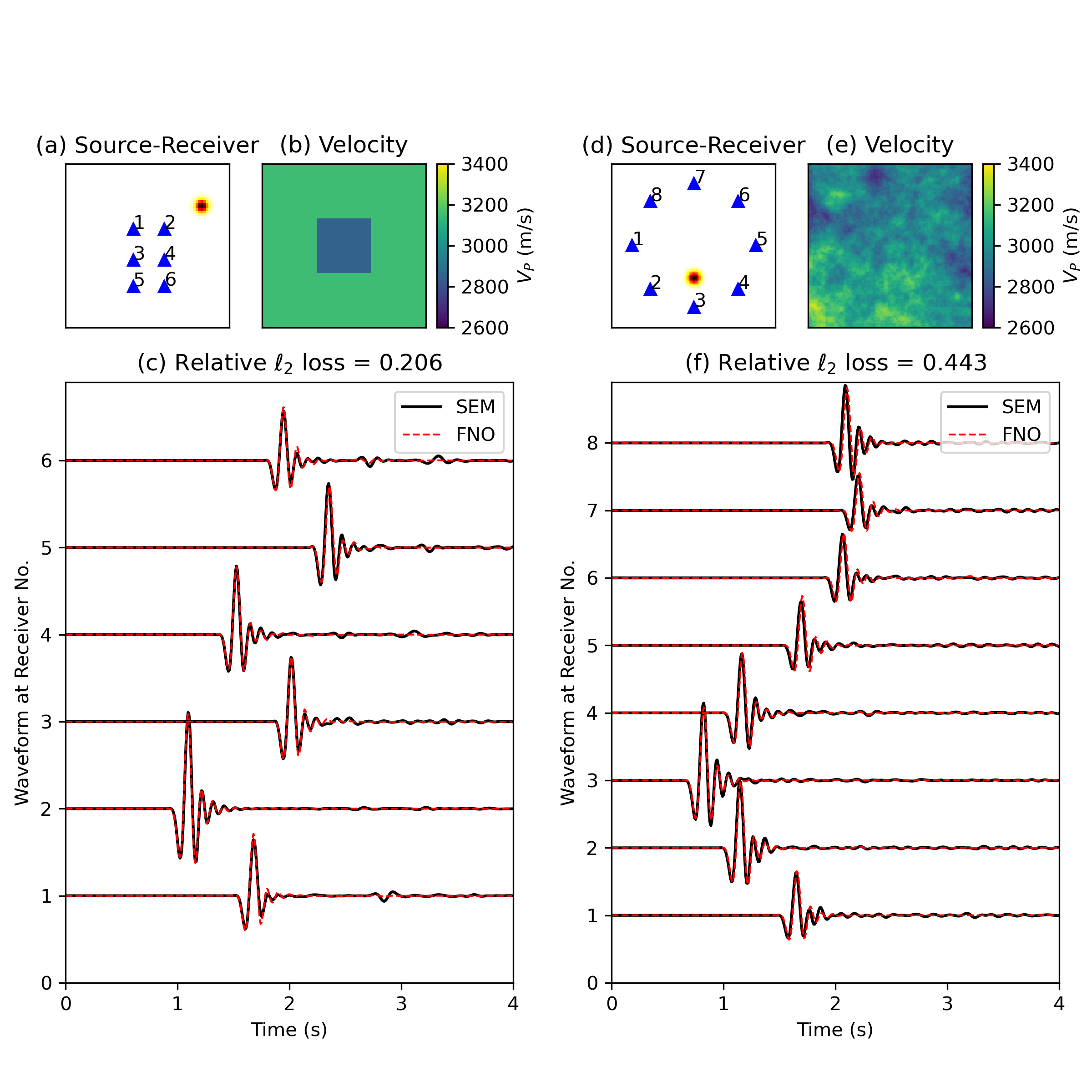}
    \caption{Model generalization experiments. (a) the source-receiver locations with receivers in blue and source in red; (b) a velocity model with a homogeneous background of 3 km/s and a 5\% square anomaly;  (c) waveform simulated with SEM (black) and FNO (red). (d-f) same as (a-c), but for an input velocity model with 2x finer resolution than trained on. These experiments show that the model is not just memorizing the solutions, but is able to generalize to entirely new conditions.}
    \label{fig:generalization}
\end{figure}

The FNO was trained only on velocity models drawn from Gaussian random fields, and while this family of functions is broad, it does not include some types of functions that exist in the Earth, such as discontinuous functions. This rises the question of whether the FNO can still generalize well to these cases. Figure \ref{fig:generalization}a-c shows an example of a predicted wavefield for a velocity model containing a constant velocity square embedded within a homogeneous medium. While the velocity model itself is rather simple, it is actually very far removed from the characteristics of the random fields that the FNO was trained on and represents a challenging example. We can see that the predicted wavefield does a very good job of approximating the wavefield compared to the ground truth. We believe the small residual errors can be reduced with better hyperparameter selection. 

\subsubsection*{ Generalization to higher resolution grids}
FNO can be viewed in some sense as a method for learning generalized Green's functions valid for arbitrary boundary conditions. Since it is intrinsically learning a mapping between function spaces, the FNO is theoretically independent of the resolution at which the functions are discretized (this is only a requirement for evaluation on a computer). One important advantage of this is that the FNO can be trained on velocity models with a certain grid spacing, and then be evaluated on velocity models with a different grid spacing at inference time. Here, we are not simply talking about interpolating the wavefield after solving the PDE; but rather, the solutions to the PDE can actually be evaluated on a higher resolution velocity model with negligible extra compute cost. To demonstrate this, Figure \ref{fig:generalization}d-f shows the FNO prediction for a random velocity field with 2x higher resolution (128$\times$128) than the models used during training, alongside the spectral element solution. The FNO solution closely approximates the spectral element solution. Note that the velocity models with different meshes have the same roughness as the training data set. Resolving more rough structure with denser spacing can be achieved by training with many more GRFs with varying correlation length scales and variance.

\subsubsection*{Full waveform inversion with Neural Operators}
One of the most useful applications of wavefield simulations is in inversion, to image the Earth's interior. The adjoint-state method is an approach to efficiently compute the gradients of an objective function with respect to parameters of interest and can be used for seismic tomography \citep[e.g.][]{tape_adjoint_2009,gebraad_bayesian_2020}. Neural Operators are differentiable by design, which enables gradient computation with reverse-mode automatic differentiation. Automatic differentiation has been shown to be
mathematically equivalent to the adjoint-state method \citep{zhu2021general}. This allows for the gradients of the wavefield to be determined with respect to the inputs (velocity model and source location). 

\begin{figure}[!ht]
    \centering
    \includegraphics[width=0.85\textwidth]{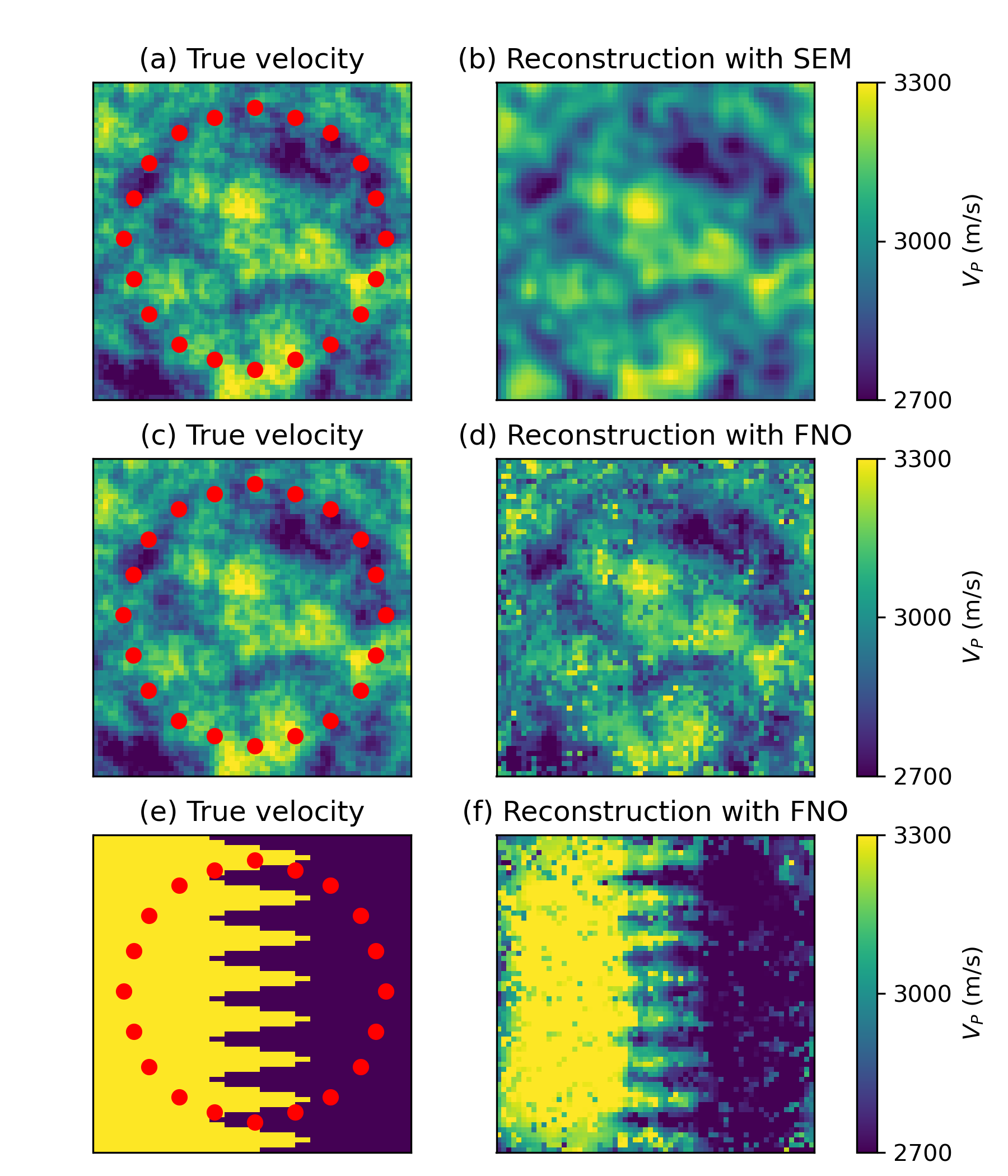}
    \caption{Example of a full waveform inversion using FNO. (a)(c)(e) True velocity models with source locations indicated by red circles and receivers placed at every node of the 64$\times$64 grid (10 km $\times$ 10 km region). (b) Reconstruction using SEM and adjoint tomography. (d)(f) Reconstruction using FNO as the forward model and automatic differentiation to compute gradients. No regularization is used for these experiments. }
    \label{fig:tomo}
\end{figure}

Figure \ref{fig:tomo} demonstrates our FWI performance. For each case, we compute synthetic observations using the source distribution as shown (red circles), taking every point in the 64x64 grid as a receiver. The observations are noise-free for this experiment. Then, we perform tomography by starting with a homogeneous initial velocity model and forward propagating a wavefield with the FNO for each source. We calculate the loss $L = \sum_i \sum_j [u^{obs} (x_i, x_j) - u^{pred} (x_i, x_j)]^2$, and compute $\nabla L$ with automatic differentiation. The velocity model is then iteratively updated with gradient descent for 1000 iterations using the Adam optimizer \citep{kingma_adam:_2014} and a learning rate of 0.01. For comparison, Fig. \ref{fig:tomo}ab shows the imaging result using SEM and adjoint-state method, with a relative $\ell_2$ misfit between the inverted and true velocity model of 0.0289.  Fig. \ref{fig:tomo}cd shows the result for the same velocity structure using FNO and automatic differentiation, with a misfit of 0.0319. Fig. \ref{fig:tomo}ef is designed to demonstrate sharp discontinuous changes with a short wavelength. The results demonstrate the remarkable capabilities of FNO to learn a general solution operator.

We note that our FWI approach does not require an adjoint wavefield to be computed, nor a cross-correlation; the gradients can be rapidly computed with GPUs using automatic differentiation. The rapid simulation makes it substantially more efficient than adjoint methods. For these experiment, 20 sources take $\sim1$ second for one tomographic iteration including the costs of computing the forward model, while the spectral element method with adjoint methods takes $\sim100$ seconds for one tomographic iteration. These time measurements are from using only a single NVIDIA Tesla V100 GPU.


\section{Discussion}

This study presents a prototype framework for applying Neural Operators to the 2D acoustic wave equation. We anticipate that the general framework would also be suitable for the 3D elastic wave equation with relatively little modification. Indeed the FNO method was applied successfully to the Navier-Stokes equations \citep{li_fourier_2021}, which can be more challenging to solve than the elastic wave equation. In our tests, we found that only a few thousand simulations were needed to train a FNO model, and from there, required negligible time to compute a new solution. Since FNO can be trained on lower resolution simulations and then generalize to higher resolution solutions once trained, this results in substantially faster computations than using traditional numerical methods at the full resolution.

One of the limitations of the approach is that the solutions are approximate, as seen in several of the figures. However, since this is a learning-based approach, the performance can be improved in the future by using a better model architecture, thorough tuning of hyperparameters, improving the size of the training dataset, using a more appropriate objective function, and various other factors. Additionally, as new developments within machine learning emerge in this area, they would be able to be incorporated. Thus, these performance metrics should only be viewed as a starting point. For some applications, the error may be enough of an issue and traditional numerical methods may be preferable; however for many other situations in geophysics, a reasonably accurate solution may be acceptable.

Among the most exciting benefits of our approach is that by training the FNO on random velocity models, the FNO is able to produce solutions for arbitrary velocity models. This is because FNO learns a general solution operator to the PDE, and not specifically the velocity model. This means that the model does not need to be retrained for each region. Thus, the approach offers the potential for a single FNO model to be used by the entire seismology community for any region of a similar size. While the initial cost of training a FNO and performing the training simulations may be expensive, it only needs to be done a single time for the community as a whole.

\section*{Acknowledgements}
The authors thank Jack Muir for helpful comments on an early version of the manuscript.

\bibliography{Arxiv}

\begin{thebibliography}{24}
\providecommand{\natexlab}[1]{#1}
\providecommand{\url}[1]{\texttt{#1}}
\expandafter\ifx\csname urlstyle\endcsname\relax
  \providecommand{\doi}[1]{doi: #1}\else
  \providecommand{\doi}{doi: \begingroup \urlstyle{rm}\Url}\fi

\bibitem[Afanasiev et~al.(2019)Afanasiev, Boehm, van Driel, Krischer, Rietmann,
  May, Knepley, and Fichtner]{afanasiev_modular_2019}
Michael Afanasiev, Christian Boehm, Martin van Driel, Lion Krischer, Max
  Rietmann, Dave~A May, Matthew~G Knepley, and Andreas Fichtner.
\newblock Modular and flexible spectral-element waveform modelling in two and
  three dimensions.
\newblock \emph{Geophysical Journal International}, 216\penalty0 (3):\penalty0
  1675--1692, March 2019.
\newblock ISSN 0956-540X.
\newblock \doi{10.1093/gji/ggy469}.
\newblock URL \url{https://doi.org/10.1093/gji/ggy469}.

\bibitem[Duputel et~al.(2015)Duputel, Jiang, Jolivet, Simons, Rivera, Ampuero,
  Riel, Owen, Moore, Samsonov, Culaciati, and Minson]{duputel_iquique_2015}
Z.~Duputel, J.~Jiang, R.~Jolivet, M.~Simons, L.~Rivera, J.-P. Ampuero, B.~Riel,
  S.~E. Owen, A.~W. Moore, S.~V. Samsonov, F.~Ortega Culaciati, and S.~E.
  Minson.
\newblock The {Iquique} earthquake sequence of {April} 2014: {Bayesian}
  modeling accounting for prediction uncertainty.
\newblock \emph{Geophysical Research Letters}, 42\penalty0 (19):\penalty0
  7949--7957, 2015.
\newblock ISSN 1944-8007.
\newblock \doi{10.1002/2015GL065402}.
\newblock URL
  \url{https://agupubs.onlinelibrary.wiley.com/doi/abs/10.1002/2015GL065402}.

\bibitem[Fichtner et~al.(2009)Fichtner, Kennett, Igel, and
  Bunge]{fichtner_full_2009}
Andreas Fichtner, Brian L.~N. Kennett, Heiner Igel, and Hans-Peter Bunge.
\newblock Full seismic waveform tomography for upper-mantle structure in the
  {Australasian} region using adjoint methods.
\newblock \emph{Geophysical Journal International}, 179\penalty0 (3):\penalty0
  1703--1725, December 2009.
\newblock ISSN 0956-540X.
\newblock \doi{10.1111/j.1365-246X.2009.04368.x}.
\newblock URL \url{https://doi.org/10.1111/j.1365-246X.2009.04368.x}.

\bibitem[Gebraad et~al.(2020)Gebraad, Boehm, and
  Fichtner]{gebraad_bayesian_2020}
Lars Gebraad, Christian Boehm, and Andreas Fichtner.
\newblock Bayesian {Elastic} {Full}-{Waveform} {Inversion} {Using}
  {Hamiltonian} {Monte} {Carlo}.
\newblock \emph{Journal of Geophysical Research: Solid Earth}, 125\penalty0
  (3):\penalty0 e2019JB018428, 2020.
\newblock ISSN 2169-9356.
\newblock \doi{10.1029/2019JB018428}.
\newblock URL
  \url{https://agupubs.onlinelibrary.wiley.com/doi/abs/10.1029/2019JB018428}.
\newblock \_eprint:
  https://agupubs.onlinelibrary.wiley.com/doi/pdf/10.1029/2019JB018428.

\bibitem[Graves and Pitarka(2016)]{graves_kinematic_2016}
Robert Graves and Arben Pitarka.
\newblock Kinematic {Ground}‐{Motion} {Simulations} on {Rough} {Faults}
  {Including} {Effects} of {3D} {Stochastic} {Velocity} {Perturbations}.
\newblock \emph{Bulletin of the Seismological Society of America}, 106\penalty0
  (5):\penalty0 2136--2153, October 2016.
\newblock ISSN 0037-1106.
\newblock \doi{10.1785/0120160088}.
\newblock URL
  \url{https://pubs.geoscienceworld.org/ssa/bssa/article/106/5/2136/350935/Kinematic-Ground-Motion-Simulations-on-Rough}.
\newblock Publisher: GeoScienceWorld.

\bibitem[Kingma and Ba(2014)]{kingma_adam:_2014}
Diederik~P. Kingma and Jimmy Ba.
\newblock Adam: {A} {Method} for {Stochastic} {Optimization}.
\newblock \emph{arXiv:1412.6980 [cs]}, December 2014.
\newblock URL \url{http://arxiv.org/abs/1412.6980}.
\newblock arXiv: 1412.6980.

\bibitem[Lee et~al.(2014)Lee, Chen, Jordan, Maechling, Denolle, and
  Beroza]{lee_full-3-d_2014}
En-Jui Lee, Po~Chen, Thomas~H. Jordan, Phillip~B. Maechling, Marine A.~M.
  Denolle, and Gregory~C. Beroza.
\newblock Full-3-{D} tomography for crustal structure in {Southern}
  {California} based on the scattering-integral and the adjoint-wavefield
  methods.
\newblock \emph{Journal of Geophysical Research: Solid Earth}, 119\penalty0
  (8):\penalty0 6421--6451, 2014.
\newblock ISSN 2169-9356.
\newblock \doi{10.1002/2014JB011346}.
\newblock URL
  \url{https://agupubs.onlinelibrary.wiley.com/doi/abs/10.1002/2014JB011346}.
\newblock \_eprint:
  https://agupubs.onlinelibrary.wiley.com/doi/pdf/10.1002/2014JB011346.

\bibitem[Li et~al.(2020{\natexlab{a}})Li, Kovachki, Azizzadenesheli, Liu,
  Bhattacharya, Stuart, and Anandkumar]{li_multipole_2020}
Zongyi Li, Nikola Kovachki, Kamyar Azizzadenesheli, Burigede Liu, Kaushik
  Bhattacharya, Andrew Stuart, and Anima Anandkumar.
\newblock Multipole {Graph} {Neural} {Operator} for {Parametric} {Partial}
  {Differential} {Equations}.
\newblock \emph{arXiv:2006.09535 [cs, math, stat]}, October 2020{\natexlab{a}}.
\newblock URL \url{http://arxiv.org/abs/2006.09535}.
\newblock arXiv: 2006.09535.

\bibitem[Li et~al.(2020{\natexlab{b}})Li, Kovachki, Azizzadenesheli, Liu,
  Bhattacharya, Stuart, and Anandkumar]{li_neural_2020}
Zongyi Li, Nikola Kovachki, Kamyar Azizzadenesheli, Burigede Liu, Kaushik
  Bhattacharya, Andrew Stuart, and Anima Anandkumar.
\newblock Neural {Operator}: {Graph} {Kernel} {Network} for {Partial}
  {Differential} {Equations}.
\newblock \emph{arXiv:2003.03485 [cs, math, stat]}, March 2020{\natexlab{b}}.
\newblock URL \url{http://arxiv.org/abs/2003.03485}.
\newblock arXiv: 2003.03485.

\bibitem[Li et~al.(2021)Li, Kovachki, Azizzadenesheli, Liu, Bhattacharya,
  Stuart, and Anandkumar]{li_fourier_2021}
Zongyi Li, Nikola Kovachki, Kamyar Azizzadenesheli, Burigede Liu, Kaushik
  Bhattacharya, Andrew Stuart, and Anima Anandkumar.
\newblock Fourier {Neural} {Operator} for {Parametric} {Partial} {Differential}
  {Equations}.
\newblock \emph{arXiv:2010.08895 [cs, math]}, May 2021.
\newblock URL \url{http://arxiv.org/abs/2010.08895}.
\newblock arXiv: 2010.08895.

\bibitem[Moseley et~al.(2020{\natexlab{a}})Moseley, Markham, and
  Nissen-Meyer]{moseley_solving_2020}
Ben Moseley, Andrew Markham, and Tarje Nissen-Meyer.
\newblock Solving the wave equation with physics-informed deep learning.
\newblock \emph{arXiv:2006.11894 [physics]}, June 2020{\natexlab{a}}.
\newblock URL \url{http://arxiv.org/abs/2006.11894}.
\newblock arXiv: 2006.11894.

\bibitem[Moseley et~al.(2020{\natexlab{b}})Moseley, Nissen-Meyer, and
  Markham]{moseley_deep_2020}
Ben Moseley, Tarje Nissen-Meyer, and Andrew Markham.
\newblock Deep learning for fast simulation of seismic waves in complex media.
\newblock \emph{Solid Earth}, 11\penalty0 (4):\penalty0 1527--1549, August
  2020{\natexlab{b}}.
\newblock ISSN 1869-9510.
\newblock \doi{10.5194/se-11-1527-2020}.
\newblock URL \url{https://se.copernicus.org/articles/11/1527/2020/}.
\newblock Publisher: Copernicus GmbH.

\bibitem[Moseley et~al.(2021)Moseley, Markham, and
  Nissen-Meyer]{moseley_finite_2021}
Ben Moseley, Andrew Markham, and Tarje Nissen-Meyer.
\newblock Finite {Basis} {Physics}-{Informed} {Neural} {Networks} ({FBPINNs}):
  a scalable domain decomposition approach for solving differential equations.
\newblock \emph{arXiv:2107.07871 [physics]}, July 2021.
\newblock URL \url{http://arxiv.org/abs/2107.07871}.
\newblock arXiv: 2107.07871.

\bibitem[Rodgers et~al.(2019)Rodgers, Petersson, Pitarka, McCallen, Sjogreen,
  and Abrahamson]{rodgers_broadband_2019}
Arthur~J. Rodgers, N.~Anders Petersson, Arben Pitarka, David~B. McCallen, Bjorn
  Sjogreen, and Norman Abrahamson.
\newblock Broadband (0–5 {Hz}) {Fully} {Deterministic} {3D}
  {Ground}‐{Motion} {Simulations} of a {Magnitude} 7.0 {Hayward} {Fault}
  {Earthquake}: {Comparison} with {Empirical} {Ground}‐{Motion} {Models} and
  {3D} {Path} and {Site} {Effects} from {Source} {Normalized} {Intensities}.
\newblock \emph{Seismological Research Letters}, 90\penalty0 (3):\penalty0
  1268--1284, May 2019.
\newblock ISSN 0895-0695.
\newblock \doi{10.1785/0220180261}.
\newblock URL
  \url{https://pubs.geoscienceworld.org/ssa/srl/article/90/3/1268/568984/Broadband-0-5-Hz-Fully-Deterministic-3D-Ground}.
\newblock Publisher: GeoScienceWorld.

\bibitem[Smith et~al.(2020)Smith, Azizzadenesheli, and
  Ross]{smith_eikonet_2020}
Jonathan~D. Smith, Kamyar Azizzadenesheli, and Zachary~E. Ross.
\newblock {EikoNet}: {Solving} the {Eikonal} {Equation} {With} {Deep} {Neural}
  {Networks}.
\newblock \emph{IEEE Transactions on Geoscience and Remote Sensing}, pages
  1--12, 2020.
\newblock ISSN 1558-0644.
\newblock \doi{10.1109/TGRS.2020.3039165}.
\newblock Conference Name: IEEE Transactions on Geoscience and Remote Sensing.

\bibitem[Smith et~al.(2021)Smith, Ross, Azizzadenesheli, and
  Muir]{smith_hyposvi_2021}
Jonathan~D. Smith, Zachary~E. Ross, Kamyar Azizzadenesheli, and Jack~B. Muir.
\newblock {HypoSVI}: {Hypocenter} inversion with {Stein} variational inference
  and {Physics} {Informed} {Neural} {Networks}.
\newblock \emph{arXiv:2101.03271 [physics]}, January 2021.
\newblock URL \url{http://arxiv.org/abs/2101.03271}.
\newblock arXiv: 2101.03271.

\bibitem[Tape et~al.(2009)Tape, Liu, Maggi, and Tromp]{tape_adjoint_2009}
Carl Tape, Qinya Liu, Alessia Maggi, and Jeroen Tromp.
\newblock Adjoint {Tomography} of the {Southern} {California} {Crust}.
\newblock \emph{Science}, 325\penalty0 (5943):\penalty0 988--992, August 2009.
\newblock ISSN 0036-8075, 1095-9203.
\newblock \doi{10.1126/science.1175298}.
\newblock URL \url{https://science.sciencemag.org/content/325/5943/988}.
\newblock Publisher: American Association for the Advancement of Science
  Section: Report.

\bibitem[Virieux and Operto(2009)]{virieux_overview_2009}
J.~Virieux and S.~Operto.
\newblock An overview of full-waveform inversion in exploration geophysics.
\newblock \emph{GEOPHYSICS}, 74\penalty0 (6):\penalty0 WCC1--WCC26, November
  2009.
\newblock ISSN 0016-8033.
\newblock \doi{10.1190/1.3238367}.
\newblock URL \url{https://library.seg.org/doi/full/10.1190/1.3238367}.
\newblock Publisher: Society of Exploration Geophysicists.

\bibitem[Wang and Zhan(2020)]{wang_moving_2020}
Xin Wang and Zhongwen Zhan.
\newblock Moving from 1-{D} to 3-{D} velocity model: automated waveform-based
  earthquake moment tensor inversion in the {Los} {Angeles} region.
\newblock \emph{Geophysical Journal International}, 220\penalty0 (1):\penalty0
  218--234, January 2020.
\newblock ISSN 0956-540X.
\newblock \doi{10.1093/gji/ggz435}.
\newblock URL \url{https://doi.org/10.1093/gji/ggz435}.

\bibitem[Xiao et~al.(2021)Xiao, Deng, and Wang]{xiao_deep-learning-based_2021}
Cong Xiao, Ya~Deng, and Guangdong Wang.
\newblock Deep-{Learning}-{Based} {Adjoint} {State} {Method}: {Methodology} and
  {Preliminary} {Application} to {Inverse} {Modeling}.
\newblock \emph{Water Resources Research}, 57\penalty0 (2):\penalty0
  e2020WR027400, 2021.
\newblock ISSN 1944-7973.
\newblock \doi{10.1029/2020WR027400}.
\newblock URL
  \url{https://agupubs.onlinelibrary.wiley.com/doi/abs/10.1029/2020WR027400}.
\newblock \_eprint:
  https://agupubs.onlinelibrary.wiley.com/doi/pdf/10.1029/2020WR027400.

\bibitem[Ye et~al.(2016)Ye, Lay, Kanamori, and Rivera]{ye_rupture_2016}
Lingling Ye, Thorne Lay, Hiroo Kanamori, and Luis Rivera.
\newblock Rupture characteristics of major and great mw $\geq$ 7.0 megathrust
  earthquakes from 1990 to 2015: 2. {Depth} dependence.
\newblock \emph{Journal of Geophysical Research: Solid Earth}, 121\penalty0
  (2):\penalty0 2015JB012427, February 2016.
\newblock ISSN 2169-9356.
\newblock \doi{10.1002/2015JB012427}.
\newblock URL
  \url{http://onlinelibrary.wiley.com/doi/10.1002/2015JB012427/abstract}.

\bibitem[Zhang and Gao(2021)]{zhang_deep-learning_2021}
Wei Zhang and Jinghuai Gao.
\newblock Deep-{Learning} {Full}-{Waveform} {Inversion} {Using} {Seismic}
  {Migration} {Images}.
\newblock \emph{IEEE Transactions on Geoscience and Remote Sensing}, pages
  1--18, 2021.
\newblock ISSN 1558-0644.
\newblock \doi{10.1109/TGRS.2021.3062688}.
\newblock Conference Name: IEEE Transactions on Geoscience and Remote Sensing.

\bibitem[Zhu et~al.(2020)Zhu, Xu, Darve, Biondi, and
  Beroza]{zhu_integrating_2020}
Weiqiang Zhu, Kailai Xu, Eric Darve, Biondo Biondi, and Gregory~C. Beroza.
\newblock Integrating {Deep} {Neural} {Networks} with {Full}-waveform
  {Inversion}: {Reparametrization}, {Regularization}, and {Uncertainty}
  {Quantification}.
\newblock \emph{arXiv:2012.11149 [physics]}, December 2020.
\newblock URL \url{http://arxiv.org/abs/2012.11149}.
\newblock arXiv: 2012.11149.

\bibitem[Zhu et~al.(2021)Zhu, Xu, Darve, and Beroza]{zhu2021general}
Weiqiang Zhu, Kailai Xu, Eric Darve, and Gregory~C Beroza.
\newblock A general approach to seismic inversion with automatic
  differentiation.
\newblock \emph{Computers \& Geosciences}, 151:\penalty0 104751, 2021.

\end{thebibliography}
\bibliographystyle{plainnat}

\end{document}